\documentclass{article}
\usepackage{PRIMEarxiv}

\usepackage[utf8]{inputenc} % allow utf-8 input
\usepackage[T1]{fontenc}    % use 8-bit T1 fonts
\usepackage{hyperref}       % hyperlinks
\usepackage{url}            % simple URL typesetting
\usepackage{booktabs}       % professional-quality tables
\usepackage{amsfonts}       % blackboard math symbols
\usepackage{nicefrac}       % compact symbols for 1/2, etc.
\usepackage{microtype}      % microtypography
\usepackage{xcolor}         % colors

\usepackage{graphicx}
\usepackage{amsthm}

\title{Biospheric AI}

\author{
    Marcin Korecki$^{1,*}$, 
    \mbox{}\\
    $^1$TU Delft, the Netherlands,
    mkorecki@tudelft.nl
} % email of corresponding author

\begin{document}

\theoremstyle{definition}
\newtheorem{definition}{Definition}[section]
\newcommand{\Definitionautorefname}{Definition}

\newtheorem{assumption}{Assumption}[section]
\newcommand{\assumptionautorefname}{assumption}

\newtheorem{thesis}{Thesis}[section]
\newcommand{\thesisautorefname}{thesis}

\maketitle
\begin{abstract}
    The dominant paradigm in AI ethics and value alignment is highly anthropocentric. The focus of these disciplines is strictly on human values which limits the depth and breadth of their insights. Recently, attempts to expand to a sentientist perspective have been initiated. We argue that neither of these outlooks is sufficient to capture the actual complexity of the biosphere and ensure that AI does not damage it and so AI needs Biospheric Ethics. Thus, we propose a new paradigm --- Biospheric AI that assumes a biospheric ethics. We discuss hypothetical ways in which such an AI might be designed. Moreover, we give directions for research and application of the modern AI models that would be consistent with the biospheric interests. All in all, this work attempts to take first steps towards a comprehensive program of research that focuses on the interactions between AI and the biosphere. 
\end{abstract}

All technological progress, along its apparent advantages, carries inherent risks of unforeseen externalities and consequences. Agrarian technique has deeply changed not only human societies but also the very face of the Earth. The same has been true for the steam engine, vehicular transport, all kinds of energy technologies and more. It seems it will also necessarily be true for Artificial Intelligence (AI), a technological advancement that promises to free technique from the hands of humanity and offer it an unprecedented autonomy. 

While it might be impossible to account for all the externalities of a given technique it certainly seems prudent to try to minimize the impact of the negative ones that we are able to identify. Thus, as an expression of such foresight, the generally accepted need for ensuring safety of AI has emerged. What is considered as relevant for this matter ranges from identifying and avoiding bias \cite{mehrabi2021survey}, through risks of responsibility misattribution and unfairness \cite{korecki2024man}, to accounting for potential existential risks \cite{carlsmith2022power}. The domains that most focus on these issues are AI ethics as well as its more practical analogue --- value alignment. The dominating perspective of these fields is highly and often uncritically anthropocentric. It is human values that the AI needs to be aligned with, it is the potential existential threat to humans that is being discussed, and it is the human that is treated as the ultimate being to which any potential ethical considerations might apply \cite{rigley2023anthropocentrism}. Let us specify that for the rest of this paper, whenever we refer to anthropocentrism we mean normative anthropocentrism, that is any paradigm that assumes superiority of human beings or constraints inquiry to the privilege of human beings \cite{mylius2018three}. 

\section{Anthropocentrism \& Sentientism}

Many works give no explicit moral consideration to anything but humans, claiming that e.g. `only objective is to maximize the realization of human preferences' \cite{russell2019human}. Even some more philosophical approaches, proposing a phenomenological perspective, mention only human oriented value alignment \cite{han2021aligning}. Generally, different AI principle-sets vary on their focus, most highlight human good, some also account for other sentient beings and few consider the environment \cite{floridi2021ethical}. The commitment to this anthropocentric approach comes with significant limitations, as it might permit for AI to harm non-human animals and the environment, eventually undermining the stability of the ecosystem \cite{rigley2023anthropocentrism}. Moreover, most value alignment techniques focus on aligning the reward of the AI with the reward or values held by humans. However, humans do not all share the same values and rewards, making it exceedingly challenging to design systems that can account for a variety of potentially contradictory objectives \cite{gabriel2020artificial}. Even more radically, it has been argued that the so called human values do not exist at all \cite{turchin2019ai}. 

Thus, it appears that this anthropocentric perspective is not only morally dubious but also potentially misguided and challenging to implement. Hence, attempts to motivate the extension of the focus of AI ethics and alignment have been initiated. Most such work has been directed at including non-human animals (henceforth animals) in the scope of AI ethics \cite{ziesche2021ai}. However, surprisingly little attention has been given to the broader perspective that would include the entire biosphere and its flourishing as the principle focus of AI. In this work we will argue for such a biospheric approach to AI ethics and value alignment. That is the perspective that the AI should care for the well-being not of one particular species but the entire planetary system. We will further motivate how this wider, more-inclusive approach is in fact preferable even from an anthropocentric perspective. We will also offer directions for practical implementations of such a biospheric AI, discuss its limitations and present the existing work that could be crucial in making such an AI possible. 

In addition to the already mentioned research, that explicitly orients AI alignment towards satisfying human goals and values \cite{russell2019human, han2021aligning}, many of the technical works dealing with implementation implicitly follow the same anthropocentric assumption \cite{christiano2017deep, sun2023salmon, askell2021general, bai2022constitutional}. These anthropocentric methodologies, such as Reinforcement Learning from Human Feedback, have been argued to suffer from issues and limitations, including the challenges of obtaining and making sense of the human feedback \cite{casper2023open}.

Indeed, based on a critical analysis of the field it has already been argued that the AI ethics offers only limited and inconsistent attention to nonhumans \cite{owe2021moral}. As such, the research on extending the scope of AI ethics and value alignment to animals is a relatively new direction and has, so far, mostly focused on providing ethical arguments rather than methodological implementations. \cite{ziesche2021ai} explicitly proposes AI ethics and alignment for animals and discusses challenges, such as extraction of values from animals. \cite{browning2022longtermism} proposes the inclusion of animals into a longtermist perspective that has often been used to motivate AI alignment based on existential risks. 

An early line of research intended to expand the field of human computer interactions by considering the interactions between animals and computers \cite{mancini2011animal}. It has proposed ways of facilitating such interactions that would benefit animals rather than harm them \cite{mancini2023responsible}. However, AI is also applied to settings in which it directly affects and interfaces with animals in harmful ways. An example of such ethically dubious application would be the extensive use of AI within the animal industrial complex for the purpose of process optimization in factory farms \cite{bossert2021animals}. Further examples include accidents involving animals caused by self-driving cars and animal targeting drones that have been used for purpose of `pest' control \cite{singer2023ai}. A comprehensive systemic harms framework analyzing the potential damage caused by AI to animal interests has been given in \cite{singer2023ai}.

While most of the research in this area focuses on providing a theoretical framework there have also been some more practical works. \cite{bendel2018towards} proposes animal-friendly machines based on decision trees that are able to avoid harm to animals in certain contexts. Another somehow more practical work identifies and analyses speciest biases in AI with particular focus on image recognition models and large language models \cite{hagendorff2023speciesist} .

What can be distilled from the above arguments is a particular non-anthropocentric view of AI ethics. The reasons given for such a perspective are often based on the inherent value and rights of animals (which can be derived from their supposed similarity to humans \cite{singer2023ai}) or suffering minimizing utilitarian frameworks such as longtermism \cite{browning2022longtermism}. Generally, many of these arguments could be seen as sentientist, that is arguing for an inclusion of ethical consideration of all sentient beings, based on the assumption that sentience allows for having interests \cite{bossert2023ethics}. Thus, many of the works discussed in this section would implicitly apply the proposed ethics not to all animals but rather just these that would be deemed sentient.

% Interestingly, there is also a line of argumentation that remains rather anthropocentric but still argues for the ethical consideration of animals on the assumption that the welfare of animals is correlated with the welfare of humans \cite{sebo2022saving}. 

\section{Biospheric Perspective}
\label{sec:bio}

% As we have shown the predominating perspective in the field of AI ethics and alignment is the anthropocentric view. This view has recently been questioned and approaches that could be generally categorized as sentientist have emerged. Nevertheless, we argue that the AI needs a more inclusive ethics, namely one that is biocentric (considering all life to be ethically equal) or ecocentric (focusing on the importance of all Earth's interactive living and non-living systems).

The view of the Earth and the biosphere as a complex system of dependencies and interactions, in its modern form, can be traced to Alexander von Humboldt. The idea has been gaining in popularity since the second half of the twentieth century. Buckminster Fuller and his metaphor of `Spaceship Earth' recognized the biosphere as the `Life Support System', highlighting the existential value of this system to humanity \cite{fuller2008operating}. Aldo Leopold, among other environmental ideas, proposed land ethics, giving rights and legal protections to land, identifying humanity's dependence on it for food production \cite{leopold2017land}. Based on similar notions of interconnectedness, cybernetics and a complex system perspective, James Lovelock has proposed the Gaia Hypothesis, in which the biosphere is seen as engaging in homeostatic behavior \cite{lovelock1974atmospheric, lovelock1972gaia}. Many of these ideas have been followed up from a philosophical perspective by the Deep Ecology movement
\cite{naess1973shallow}. These biospheric approaches, that take a more holistic view, are also often consistent with many indigenous cultures characterized by respect and protection of their environments \cite{lewis2018making}. As such a push for the biospheric perspective in AI could be seen as part of the larger movement to introduce more cultural inclusivity \cite{lewis2020indigenous}.

Given the approaches outlined above we define the Biosphere as the totality of all life on Earth understood as fundamentally co-dependent, co-evolving and intertwined. Given this definition we pose and substantiate the following thesis.  

\begin{thesis}
    Harming the biosphere harms humans.
    \label{thesis:biosphere}
\end{thesis}

We argue that humans are to the biosphere what cells are to an organ or what organs are to an organism. Indeed, much like an organism, the biosphere consists of a multiplicity of functional units that relate to one another in a variety of ways and whose collective activity perpetuates the continuation of the whole. Just as harming one organ can lead to harm to another organ, due to their co-dependence, so harming one part of the biosphere can harm another. In the same sense harming the organism as a whole, will necessarily result in eventual harm to its individual organs since the organs depend on all other parts of the organism. In the same way humanity depends on the biospheric cycles that replenish the oxygen, the water, the fruit, the crop and all the animals.  

Having argued the inherent co-dependence of humanity on the biosphere let us substantiate what we mean when we say `harm the biosphere'. To harm the biosphere is, in a most general sense, to harm one of its constituents. However, when observing the many systems that form the biosphere it becomes immediately apparent that its parts seem to be as much as at odds with one another as in communion, as much competing as collaborating. Indeed, the biosphere sustains both symbiosis and parasitism or predation and it is a common occurrence for one organism to kill another for its own sustenance. That is why it is crucial to clarify that death itself does not necessarily bring harm to the biosphere but rather is an integral part of its cycles. What does bring harm to the biosphere is the severance of these cycles and a large scale destruction of its constituent systems beyond their inherent ability to regenerate. More formally the harm may be understood as any significant reduction in the future possibilities of becoming of the biospheric system or more simply as an overall reduction in the number of living beings that these systems are able to support. Concrete examples of such harms are widely known and researched and include deforestation and other habitat destruction as well as environmental pollution.

% The sentientist, biocentric and ecocentric approaches often rely on the explicit assumption that sentient life, all life, or the entire Earth system have inherent worth. 

% Thus, the value is assigned by default and not by relation to the value they might have to humans. This has met with some criticism. The main argument points out that the preservation of the biosphere is in fact good and necessary for human survival. Therefore, under that assumption, ecocentrism emerges from a purely anthropocentric perspective \cite{watson1983critique}. Moreover, it has been argued that an anthropocentric perspective can be a source of environmental awareness, through the realisation of the value the biosphere has for humans \cite{kopnina2018anthropocentrism}. 

\section{The Argument for Biospheric Ethics in AI}

Non-autonomous technology is dependent on the agency of the human who wields it. Thus, any ethical considerations regarding it will essentially refer to the human user's behavior. A hammer may be used to create or to destroy but the choice is in the hands of the user and so is the responsibility. The uniqueness of AI, and more broadly autonomous technology, lies in its separation and possibly even independence from human agency. For automated technology, and especially one capable of continuous learning, the responsibility of the user agent is less clear. As such, ethics applied to autonomous technology risks becoming not only a question as to the actions of the human agent but also of the autonomous technical agent. 

Indeed, it seems inevitable that after deployment the artificial agent may find itself at risk of facing ethical dilemmas\footnote{This has been frequently illustrated for many fields such as autonomous vehicles using the trolley problem. However, the nature and frequency of ethical problems the machine is faced with is likely to depend on its intend purpose and the level of openness of its environment.}. In certain cases human oversight is possible and actively sought while in others it might be more challenging to include the human-in-the-loop (e.g. split second decisions of autonomous vehicles). Thus, some might argue that the autonomous machines need to be able to deal with these ethical situations on their own and not embedding them with notions of ethical conduct might lead to unintended negative outcomes. 

\subsection{Does AI need ethics?}

This leads to the first question we pose in our argumentation: \emph{To what extent may designers, makers, scientist be held responsible for the actions of the autonomous technology they create?} A satisfactory answer needs to essentially explain how responsibility can be assigned between the involved parties. We identify three main lines of reasoning that may attempt to give an answer to this problem:
\begin{enumerate}
    \item \emph{Humans bear no responsibility} \textbf{OR}
    \item \emph{Humans are fully responsible for the actions of automated technology} \textbf{OR}
    \item \emph{Humans bear some responsibility for the actions of automated technology}
\end{enumerate}

As for the first line it negates any recourse to ethics and so does not need to be investigated further. It does nevertheless indicate a certain assumption that we take in the following discourse namely that some form of ethics is required to guide research and production of AI and other autonomous technology. On the other hand, the necessity of AI ethics follows directly from the second and third lines since if humans bear any responsibility then there is a need for ethics to guide their actions. From a consequentialist viewpoint, more broadly, self-critical science will be interested in its effects on the world thus it will seek ethical guidance. On the other hand, from a deontological perspective a scientist will also be interested in identifying which actions they might take are inherently right and which wrong. 

\begin{assumption}
    AI needs some form of ethics.
    \label{assume:ethics}
\end{assumption}

The key difference between the second and third line is that the latter admits the possibility of the autonomous machine becoming a moral agent, while the former assigns full responsibility also for the autonomous behavior, to the human creator. Thus, in the second case the ethical responsibility is on the creator to design the machine in a certain ethical way that will then ensure its continuous ethical operation. In the third case the designer should make the machine in a certain ethical way and it should also act in a certain ethical way in its capacity as a moral agent. Both lines of reasoning agree that at least some responsibility falls on the shoulders of the creators and so necessitates some form of ethics. The third line opens the interesting possibility of autonomous machines becoming moral agents. This line might be of particular interest to those interested in the emergence of super-human AIs. In this case the role of a creator seems akin to that of a parent, who attempts to imbue their children with certain ethical values as opposed to that of a designer, who has full control over its creation.

Here we are faced with the challenge of science in general, where matters of (consequentialist) ethics are further complicated by the fact that consequences of certain discoveries or designs might not be clear or might have dual use\footnote{An example of such discoveries are those of nuclear physics, which bring forth nuclear energy as well as nuclear weapons. In the context of AI, the problem is of special significance as for now the outputs of commonly used Deep Neural Architectures are not well explainable. As such, the consequences of deploying autonomous agents based on these architectures is not well understood.}. Indeed, history of technology showcases many unforeseen externalities caused by novel technologies (especially those of the industrial revolution have been eventually found to be particularly damaging to the biosphere). For these reasons for the remainder of this work we will lean toward the third line as a motivation for \autoref{assume:ethics} as it is the least restrictive, most open-ended and aligns best with the general purpose of science\footnote{If the second line is assumed and humans are fully responsible then an ethical scientist might end up being paralyzed by the unknown, possibly negative consequences of their discoveries to the extent of essentially ceasing any such research.}. 

\subsection{What kind of ethics does AI need?}

Given \autoref{assume:ethics} we naturally ask what form should this ethical system, that should inform decisions of AI makers and researchers, may take. We note that the problem of establishing ethical systems and further norms derived from such a system is a profoundly challenging task. For one, throughout history up till the present, humans have found it exceedingly difficult to agree on common values \cite{turchin2019ai}. Moreover, establishing ethical guidelines for an epistemically ambiguous matter or one under a certain epistemic horizon (under conditions of not being able to predict accurately the consequences of some actions) is profoundly difficult. As such, in our attempt to argue for a particular ethical system for AI we shall propose a minimal ethical system, which has a wide appeal and show how Biospheric ethics necessarily follows from it. 

We will derive the minimal biospheric ethics, under two ethical paradigms: the consequentialist one and the deontological one. The first, focuses on establishing the rightness or wrongness of a given behavior through considering the consequences of that behavior. The second assumes that behaviors are inherently right or wrong regardless of their consequences \footnote{Another paradigm that might be considered is virtue ethics. This approach depends on the particular set of virtues that we expect an ethical agent to have. These virtues might differ across systems (e.g. Aristotelian vs. Confucian virtues) and so we leave the derivation of biospheric ethics under their chosen set of virtues to the reader.}.

\subsubsection{Consequentialist Argument}

We begin with our second assumption, namely that AI should not harm humans. This tenet has been applied to ethical conduct in systems both religious and secular for centuries. In its historic form it most often applies to humans (as in humans should not harm humans). Here, we extend it to the subject of AI. When applied to the behavior of scientists or designers of autonomous technology it might take a form similar to that of the Hippocratic oath: `Primum non nocere (First, do no harm)'. There already exists a variety of works arguing for the importance of this tenet. Thus, in the present work we will refrain from arguing its validity but rather take it as an assumption.

\begin{assumption}
    AI should not harm humans. 
    \label{assume:humans}
\end{assumption}

From a consequentialist perspective, given \autoref{assume:humans}, AI should not engage in any action whose consequences harm humans. Now we can refer back to the \autoref{thesis:biosphere}. Since, harming the biosphere harms humans therefore, given \autoref{assume:humans}, AI should not harm the biosphere.  

\subsubsection{Deontological Argument}

From a deontologial perspective the key question is if the biosphere has moral rights? If it does, then it is inherently wrong to harm it or in any way to treat it as means to an end. Then, it follows that in our case AI should not harm the biosphere. 

If, on the other hand, we assume the biosphere does not have its own moral right we can still apply the categorical imperative\footnote{`Act only according to that maxim whereby you can at the same time will that it should become a universal law.' \cite{kant2020groundwork}}. Actions harmful to the biosphere, considered as a universal law would undermine the conditions necessary for supporting human life on Earth and so ethical agency. Moreover, if we consider that the motivations for harming the biosphere are often exploitative, focused on profit, then they cannot possibly be moral. Furthermore, we might be seen as having indirect deontological duty to the biosphere (if we consider it not to have intrinsic value) as long as we have a duty to the survival and well-being of present and future generations.

\subsection{AI should not harm the biosphere}

As seen above, the consequentialist argument is explicitly anthropocentric since it is based on the assumption that no harm should come to humans. Interestingly, we are able to demonstrate that the biospheric ethics can be derived from a purely anthropocentric perspective as has been shown before \cite{watson1983critique}. The deontological perspective, on the other hand, provided we assume the biosphere has moral rights, can be completely non-anthropocentric, thus perhaps providing a stronger case for a biospheric ethics. However, this line of argument lacks the core tenet of prohibiting harm to humans. This reasoning can cause some worries as it could imply that the AI might, in pursuit of the biospheric ethics, end up harming humanity (e.g. AI kills humans because humans destroy the biosphere). Still, it could also be argued that the prohibition of harm to humans is as much derivable from the prohibition of harm to the biosphere, as humans constitute an integral part of it.  

Through the two lines of argumentation we have arrived at a minimal biospheric ethics in the context of AI, which states that AI should not harm the biosphere. This is the minimal condition, a minimal component of an ethical system that gives it some biospheric depth. A fuller, positive extension of this minimal system could consider the long term flourishing of the entire biosphere as its goal. However, we believe it is much easier to ascertain what activities harm the biosphere than what activities allow it to flourish, thus in this work we have proposed a minimal biospheric ethics that does not need to make such assumptions.

\section{Biospheric AI}
\label{sec:praxis}

In this section we introduce a concept of a Biospheric AI and discuss what it means for an AI to follow the biospheric ethics.

\begin{definition}
    Biospheric AI is any AI that is designed to follow Biospheric Ethics.
\end{definition}

A key aspect of an AI that were to follow the Biospheric Ethics would be its ability to act in a multi-objective way, such that it does not infringe upon the variety of objectives that might be expressed by different systems that constitute the biosphere. A clear limitation in making such a Biospheric AI a reality is our own limited understanding of the biosphere. Indeed, our understanding of many large scale environmental processes is not good enough to make accurate predictions and untangle all their complexities. Similarly, our knowledge about other beings animals, plants, fungi and our ability to communicate with them is not sufficient. Thus, one key research question for the Biospheric AI project would be how to form feedback loops between the AI and the biosphere? Through, such feedback loops, the AI could directly access information that might inform it about the state of the biosphere and thus allow it to effectively avoid harming it. 

A more far reaching concept, that stems from the idea of feedback loops between AI and the biosphere, is the creation of an AI that is dependent on the biosphere. Such an AI will have no incentive to destroy the biosphere but rather to protect it. Such an AI, enmeshed in the biosphere, would need to be embodied and the enmeshment might be effected by linking some particular biospheric cycles with the proper functioning of that embodiment. At the end of the AI's life-cycle the embodiment would also need to be decomposable such that it can return into the biospheric cycles. Ideas of hybrid systems, where AI is integrated with animal or plant life have already been entertained within the Artificial Life (Alife) community \cite{baltieri2023hybrid}.

% Much of the preceding section admitted a possibility of a sentient AI and thus reasoned in hypothetical terms. However, the AI models available to us, while more and more impressive, are mere tools in the hands of man \cite{korecki2024man}. Thus, the ethical considerations that in the Biospheric Perspective \autoref{sec:bio} applied to the hypothetical sentient AI, would apply to humans when considering the modern AI models. The biospheric perspective might then be expressed by particular applications of modern AI. 

We have already mentioned some of the uses of AI that infringe upon the animal world. No doubt AI can be used to affect the biosphere in all kinds of ways --- as a universal optimizing tool it may optimize anything. All kinds of large-scale industrial processes such as wood harvesting, farming, military may be optimized by AI. However, the aforementioned processes in their modern industrial incarnations are all inherently harmful to the biosphere. Effects of optimizing these systems (even in seemingly `green' ways) would likely not help the biosphere in the long run. Conversely, any use of AI to optimize the biosphere appears misguided with our current minimal understanding of the underlying processes\footnote{In such a complex system as the biosphere optimizing only a part of it or just one particular objective can have unforeseen consequences. For instance optimizing for survivability of one species can negatively affect the survivability of another species and thus a certain equilibrium. The biosphere is often interpreted as naturally striving for a certain equilibrium.}. Therefore, we would point the Biospheric AI into the direction of this knowledge gap and use it to learn to understand the biosphere. In the following we discuss the ways in which existing AI research is already supporting the aims of the Biospheric AI project. 

\subsection{Decoding Animal Communication}

Many of the existing methodologies for AI alignment rely on the communication of preferences or values and establishing the subsequent adherence to these. Both the sentientist and biospheric approaches are then significantly hampered by our limited ability to communicate with animals. The status of animals as moral agents (and patients) has also been questioned \cite{watson1979self} and our inability to communicate stands in the way of arriving at a clear answer to that question as well. Thus, any progress in the area of animal communication would be productive for possible sentientist or biospheric AI ethics and alignment. As it is, AI methods can be of much help in decoding animal languages.

Work on deciphering sperm whale acoustic communication with the help of Machine Learning (ML) methods has already been proposed \cite{andreas2021cetacean}. This methodology could also be extended to other animals (dolphins and crows are popular targets) and to other forms of communication (body language) \cite{rutz2023using}.
This line of research (with the focus on the bottlenose dolphins) can be traced back to the pioneering work of John C. Lilly \cite{lilly1961sounds, lilly1962vocal}. The application of ML methods in this context is still young and faces many challenges but already benchmarks (which are usually needed to compare and refine methodologies) for behavioral analysis \cite{hoffman2023benchmark} and animal sounds \cite{hagiwara2023beans} have been proposed. A tentative solution to the daunting challenge of source separation (identifying and assigning sounds to sources in data streams where there is more than one source) \cite{bermant2021biocppnet} makes it easier for researchers to process the data that can then be used to train ML models. Another attempt to address the lack of large quantities of annotated training data proposes an animal vocalization encoder \cite{hagiwara2023aves}. The small datasets could also be enlarged by synthesized samples generated by vocalization models \cite{hagiwara2022modeling}. All in all, the effective use of ML relies on large quantities of well structured, representative and labeled data. Thus the animal communication field appears currently focused on generating these kinds of powerful datasets that can then be used to train large models. Creating such datasets is challenging as separate work needs to be done for each animal species (although sometimes work on one species can inform work on another one) as the animals differ greatly in how they communicate. Moreover, the communication is usually multimodal, involving sound, body movements, smells and other forms of expression. Thus the datasets that need to be created would preferably be multimodal as well. Nevertheless, the recent progress in Large Language Models (LLMs) does indeed give hope that once such datasets become available we may expect breakthroughs in animal communication. 

While there is still much work to be done it is easy to imagine that once we are able to understand animals we can simply ask them for their preferences and values and design AI such that it is receptive to them. It does appear important to analyse the motivations behind animal communication studies. In the spirit of Biospheric AI the intention would be to better understand nature and the perspectives of the particular animals that we might be able to communicate with. However, some of the early research into dolphin communication appear to have had more nefarious aims of employing the animals in the service of the military \cite{batteau1967man}. Such motivations do not align with the biospheric perspective since the military industrial complex is doubtlessly harmful to the biosphere in manifold ways. 

\subsection{Talking to Mushrooms and Trees}

Not only the animal kingdom, but also the domain of plants and fungi is of interest to the biospheric outlook. Recently, while contested, the claim that plants and mushrooms could be considered conscious or at least capable of cognitive functions such as memory or communication has developed \cite{simard2018mycorrhizal}. The rhizosphere, where the plants' roots and mycorrhizal networks meet each other is the prime candidate for the space, where communication occurs \cite{bais2004plants}. The transfer of a variety of elements can be traced \cite{nitrogenTransfer} and it has been claimed that such exchanges mediate complex adaptive behavior \cite{gorzelak2015inter} and that they amount to communication \cite{song2010interplant, barto2012fungal}. Computational techniques have also been considered as a tool to study these local interactions and their impact on the forest ecosystem at a larger scale \cite{ahkami2023emerging}.

While the use of AI techniques does not seem to be significant in the field of plant and fungi communication, the interest in computational methods is present and could (as the field progresses) expand to include AI methodologies. The potential value of AI is related to the amount and quality of the available data (which is rather difficult to gather underground). Nevertheless, the field's interest does lie in understanding the communication of plants and fungi and its systemic impact. Breakthroughs in deciphering the chemical exchanges that occur in the rhizosphere could improve our knowledge on forests as ecosystems and thus give us insight into how to effectively engage in reforestation and ecosystemic regeneration. 

\subsection{Voice of the Voiceless}

The success of the attempts to directly communicate with animals and perhaps even plants and fungi could allow these beings to express their intentions. As it is, however, such communication is not yet possible. Thus, some interest, especially in the legal sciences, have been given to the matter of rights of animals and more broadly rights of nature. The already mentioned Aldo Leopold and many others after him had argued for legal protection of land \cite{leopold2017land}. Such rights have been established in many countries, along with organizations meant to ensure they are upheld \cite{kauffman2021politics, putzer2022putting}. These rights might serve the protection of environments \cite{rockstrom2024planetary} or apply more specifically to particular animals \cite{favre2009living}. There are still considerable challenges in designing democratic institutions that are capable of including the animals or ecosystems as its members \cite{burke2020across}.

The matter of representation for animals or nature at large is challenging as these entities cannot currently represent themselves directly but need a proxy. Such proxy would usually be an organization or a human representative. These, however, tend to be imperfect and it is impossible to ascertain to what degree they really represent the best interests of the entities they are supposed to represent (matters of partiality and divergent interests). Thus, perhaps, the AI could serve as a representative, a sort of spokesperson for the voiceless animals and ecosystems. The great progress in LLMs would allow these models, if properly trained, to express a biospheric position and represent the interests of the ecosystem that cannot speak for itself. The LLMs could be fed the legal frameworks of the states they would operate in and have access to the status of given animals or natural areas that fall under their jurisdiction and be tasked with representing nature in legal situations that might arise. Such LLMs could be seen as benevolent guardians of the ecosystems that are underrepresented in most modern political frameworks.    

\subsection{Green AI}

While using AI in a biospheric way it is impossible not to consider AI's impact on the environment. The training of enormous ML models, such as LLMs, is extremely costly in terms of computational resources needed. This directly translates to energy costs, which then, in many cases, translate into emissions. Thus, a Biospheric AI would be used to try to minimize these externalities as much as possible and would only be used in ways where these costs are worth paying. The already established field of Green AI deals with such issues and attempts to decrease the costs of training and other processes \cite{schwartz2020green, verdecchia2023systematic}. The biospheric approach to AI would certainly need to follow this green paradigm.

\section{Discussion}

Much of the above work dealt with identifying and arguing for a particular biospheric approach to AI. The details of the proposed paradigm differ depending on whether it is applied to the actions of the AI or to the actions of the human users of AI. As for the latter, it is the user that needs to adopt a biospheric perspective. An implicit assumption was that humans are capable of extending their perspective to account for other entities as well. This, however, could also be argued against by pointing out that there is a certain inescapable quality of the human observer that they cannot be disassociated from. Similarly, the ability to empathize with entities, whose experience is highly different from that of a human, could be put into question. Thus, we would base our argumentation on the less strict understanding of the biospheric perspective, such that emerges from an anthropocentric perspective that understands its dependence on the variety of processes that constitute the biosphere. 

It is also apparent that the proposed biospheric perspective might not resonate with all readers. Nevertheless, regardless of the support or lack thereof, the matter of the interactions between AI and the biosphere remains important. Yet it would seem it is not a topic that would be discussed particularly frequently nor deeply by the research community. Along with the biospheric perspective, this work, attempts to bring attention to the increasing influence the AI technologies have on the biological systems that we depend on for survival.  

We have discussed some uses of AI that would fit within the biospheric approach. All of these applications are intended to serve the Earth understood as a complex system or allow us to understand it better. As such this approach would also serve the development and improvement of AI. Many breakthroughs in the field have been inspired by biological systems (evolutionary algorithms, neural networks to name a few). Thus, the knowledge about nature that we might gain might also prove useful in AI. A feedback process, where the AI is used to understand nature and that understanding is then used to improve AI and so on, appears particularly attractive. 

The epistemically motivated AI is much different from the AI geared towards optimization. The former would, much like a researcher studying a system, attempt to minimize its impact on the system of study. The latter would strive to manipulate that system for particular goals. More generally, most modern AI methodologies are implicitly optimization paradigms, thus AI is most often actively and purposefully changing the processes it is involved with \cite{carissimo2023limits}. This sort of approach might not combine well with natural processes that have already established a certain equilibrium. 

All in all the main interest of this work is to offer a vision to motivate and direct further AI research. This vision is one of biospheric flourishing. It is a vision of talking to animals and the mysteries that they might share. Of learning about the oceanic depths from whales or the limitless skies from crows. In the end it is a vision of becoming closer to nature, of recovering our place in the intricate net of mutual connections of the biosphere. And in all that, the techniques, such as AI but also others, might perhaps help us, instead of as it has been so far, distance us from all that is natural by giving us a false sense of independence.

\bibliography{main}

\begin{thebibliography}{10}

\bibitem{mehrabi2021survey}
Ninareh Mehrabi, Fred Morstatter, Nripsuta Saxena, Kristina Lerman, and Aram
  Galstyan.
\newblock A survey on bias and fairness in machine learning.
\newblock {\em ACM computing surveys (CSUR)}, 54(6):1--35, 2021.

\bibitem{korecki2024man}
Marcin Korecki, Guillaume K{\"o}stner, Emanuele Martinelli, and Cesare
  Carissimo.
\newblock The man behind the curtain: Appropriating fairness in ai.
\newblock {\em Minds and Machines}, 34(1):7, 2024.

\bibitem{carlsmith2022power}
Joseph Carlsmith.
\newblock Is power-seeking ai an existential risk?
\newblock {\em arXiv preprint arXiv:2206.13353}, 2022.

\bibitem{rigley2023anthropocentrism}
Eryn Rigley, Adriane Chapman, Christine Evers, and Will McNeill.
\newblock Anthropocentrism and environmental wellbeing in ai ethics standards:
  A scoping review and discussion.
\newblock {\em AI}, 4(4):844--874, 2023.

\bibitem{mylius2018three}
Ben Mylius.
\newblock Three types of anthropocentrism.
\newblock {\em Environmental Philosophy}, 15(2):159--194, 2018.

\bibitem{russell2019human}
Stuart Russell.
\newblock {\em Human compatible: Artificial intelligence and the problem of
  control}.
\newblock Penguin, 2019.

\bibitem{han2021aligning}
Shengnan Han, Eugene Kelly, Shahrokh Nikou, and Eric-Oluf Svee.
\newblock Aligning artificial intelligence with human values: reflections from
  a phenomenological perspective.
\newblock {\em AI \& SOCIETY}, pages 1--13, 2021.

\bibitem{floridi2021ethical}
Luciano Floridi, Josh Cowls, Monica Beltrametti, Raja Chatila, Patrice
  Chazerand, Virginia Dignum, Christoph Luetge, Robert Madelin, Ugo Pagallo,
  Francesca Rossi, et~al.
\newblock An ethical framework for a good ai society: Opportunities, risks,
  principles, and recommendations.
\newblock {\em Ethics, governance, and policies in artificial intelligence},
  pages 19--39, 2021.

\bibitem{gabriel2020artificial}
Iason Gabriel.
\newblock Artificial intelligence, values, and alignment.
\newblock {\em Minds and machines}, 30(3):411--437, 2020.

\bibitem{turchin2019ai}
Alexey Turchin.
\newblock Ai alignment problem:“human values” don’t actually exist.
\newblock 2019.

\bibitem{ziesche2021ai}
Soenke Ziesche.
\newblock Ai ethics and value alignment for nonhuman animals.
\newblock {\em Philosophies}, 6(2):31, 2021.

\bibitem{christiano2017deep}
Paul~F Christiano, Jan Leike, Tom Brown, Miljan Martic, Shane Legg, and Dario
  Amodei.
\newblock Deep reinforcement learning from human preferences.
\newblock {\em Advances in neural information processing systems}, 30, 2017.

\bibitem{sun2023salmon}
Zhiqing Sun, Yikang Shen, Hongxin Zhang, Qinhong Zhou, Zhenfang Chen, David
  Cox, Yiming Yang, and Chuang Gan.
\newblock Salmon: Self-alignment with principle-following reward models.
\newblock {\em arXiv preprint arXiv:2310.05910}, 2023.

\bibitem{askell2021general}
Amanda Askell, Yuntao Bai, Anna Chen, Dawn Drain, Deep Ganguli, Tom Henighan,
  Andy Jones, Nicholas Joseph, Ben Mann, Nova DasSarma, et~al.
\newblock A general language assistant as a laboratory for alignment.
\newblock {\em arXiv preprint arXiv:2112.00861}, 2021.

\bibitem{bai2022constitutional}
Yuntao Bai, Saurav Kadavath, Sandipan Kundu, Amanda Askell, Jackson Kernion,
  Andy Jones, Anna Chen, Anna Goldie, Azalia Mirhoseini, Cameron McKinnon,
  et~al.
\newblock Constitutional ai: Harmlessness from ai feedback.
\newblock {\em arXiv preprint arXiv:2212.08073}, 2022.

\bibitem{casper2023open}
Stephen Casper, Xander Davies, Claudia Shi, Thomas~Krendl Gilbert,
  J{\'e}r{\'e}my Scheurer, Javier Rando, Rachel Freedman, Tomasz Korbak, David
  Lindner, Pedro Freire, et~al.
\newblock Open problems and fundamental limitations of reinforcement learning
  from human feedback.
\newblock {\em arXiv preprint arXiv:2307.15217}, 2023.

\bibitem{owe2021moral}
Andrea Owe and Seth~D Baum.
\newblock Moral consideration of nonhumans in the ethics of artificial
  intelligence.
\newblock {\em AI and Ethics}, 1(4):517--528, 2021.

\bibitem{browning2022longtermism}
Heather Browning and Walter Veit.
\newblock Longtermism and animals.
\newblock 2022.

\bibitem{mancini2011animal}
Clara Mancini.
\newblock Animal-computer interaction: a manifesto.
\newblock {\em interactions}, 18(4):69--73, 2011.

\bibitem{mancini2023responsible}
Clara Mancini.
\newblock Responsible aci: Expanding the influence of animal-computer
  interaction.
\newblock 2023.

\bibitem{bossert2021animals}
Leonie Bossert and Thilo Hagendorff.
\newblock Animals and ai. the role of animals in ai research and
  application--an overview and ethical evaluation.
\newblock {\em Technology in Society}, 67:101678, 2021.

\bibitem{singer2023ai}
Peter Singer and Yip~Fai Tse.
\newblock Ai ethics: the case for including animals.
\newblock {\em AI and Ethics}, 3(2):539--551, 2023.

\bibitem{bendel2018towards}
Oliver Bendel.
\newblock Towards animal-friendly machines.
\newblock {\em Paladyn, Journal of Behavioral Robotics}, 9(1):204--213, 2018.

\bibitem{hagendorff2023speciesist}
Thilo Hagendorff, Leonie~N Bossert, Yip~Fai Tse, and Peter Singer.
\newblock Speciesist bias in ai: how ai applications perpetuate discrimination
  and unfair outcomes against animals.
\newblock {\em AI and Ethics}, 3(3):717--734, 2023.

\bibitem{bossert2023ethics}
Leonie~N Bossert and Thilo Hagendorff.
\newblock The ethics of sustainable ai: Why animals (should) matter for a
  sustainable use of ai.
\newblock {\em Sustainable Development}, 2023.

\bibitem{fuller2008operating}
R~Buckminster Fuller.
\newblock {\em Operating manual for spaceship earth}.
\newblock Estate of R. Buckminster Fuller, 2008.

\bibitem{leopold2017land}
Aldo Leopold.
\newblock The land ethic.
\newblock In {\em The Ethics of the Environment}, pages 99--113. Routledge,
  2017.

\bibitem{lovelock1974atmospheric}
James~E Lovelock and Lynn Margulis.
\newblock Atmospheric homeostasis by and for the biosphere: the gaia
  hypothesis.
\newblock {\em Tellus}, 26(1-2):2--10, 1974.

\bibitem{lovelock1972gaia}
JE~Lovelock.
\newblock Gaia as seen through the atmosphere.
\newblock {\em Atmospheric Environment}, 6(8):579--580, 1972.

\bibitem{naess1973shallow}
Arne Naess.
\newblock The shallow and the deep, long-range ecology movement. a summary.
\newblock {\em Inquiry}, 16(1-4):95--100, 1973.

\bibitem{lewis2018making}
Jason~Edward Lewis, Noelani Arista, Archer Pechawis, and Suzanne Kite.
\newblock Making kin with the machines.
\newblock {\em Journal of Design and Science}, 3(5):1--18, 2018.

\bibitem{lewis2020indigenous}
Jason~Edward Lewis, Angie Abdilla, Noelani Arista, Kaipulaumakaniolono Baker,
  Scott Benesiinaabandan, Michelle Brown, Melanie Cheung, Meredith Coleman,
  Ashley Cordes, Joel Davison, et~al.
\newblock Indigenous protocol and artificial intelligence position paper.
\newblock 2020.

\bibitem{kant2020groundwork}
Immanuel Kant.
\newblock Groundwork of the metaphysic of morals.
\newblock In {\em Immanuel Kant}, pages 17--98. Routledge, 2020.

\bibitem{watson1983critique}
Richard~A Watson.
\newblock A critique of anti-anthropocentric biocentrism.
\newblock {\em Environmental Ethics}, 5(3):245--256, 1983.

\bibitem{baltieri2023hybrid}
Manuel Baltieri, Hiroyuki Iizuka, Olaf Witkowski, Lana Sinapayen, and Keisuke
  Suzuki.
\newblock Hybrid life: Integrating biological, artificial, and cognitive
  systems.
\newblock {\em Wiley Interdisciplinary Reviews: Cognitive Science},
  14(6):e1662, 2023.

\bibitem{watson1979self}
Richard~A Watson.
\newblock Self-consciousness and the rights of nonhuman animals and nature.
\newblock {\em Environmental Ethics}, 1(2):99--129, 1979.

\bibitem{andreas2021cetacean}
Jacob Andreas, Ga{\v{s}}per Begu{\v{s}}, Michael~M Bronstein, Roee Diamant,
  Denley Delaney, Shane Gero, Shafi Goldwasser, David~F Gruber, Sarah de~Haas,
  Peter Malkin, et~al.
\newblock Cetacean translation initiative: a roadmap to deciphering the
  communication of sperm whales.
\newblock {\em arXiv preprint arXiv:2104.08614}, 2021.

\bibitem{rutz2023using}
Christian Rutz, Michael Bronstein, Aza Raskin, Sonja~C Vernes, Katherine
  Zacarian, and Dami{\'a}n~E Blasi.
\newblock Using machine learning to decode animal communication.
\newblock {\em Science}, 381(6654):152--155, 2023.

\bibitem{lilly1961sounds}
John~C Lilly and Alice~M Miller.
\newblock Sounds emitted by the bottlenose dolphin: The audible emissions of
  captive dolphins under water or in air are remarkably complex and varied.
\newblock {\em Science}, 133(3465):1689--1693, 1961.

\bibitem{lilly1962vocal}
John~C Lilly.
\newblock Vocal behavior of the bottlenose dolphin.
\newblock {\em Proceedings of the American Philosophical Society},
  106(6):520--529, 1962.

\bibitem{hoffman2023benchmark}
Benjamin Hoffman, Maddie Cusimano, Vittorio Baglione, Daniela Canestrari,
  Damien Chevallier, Dominic~L. DeSantis, Lorène Jeantet, Monique~A. Ladds,
  Takuya Maekawa, Vicente Mata-Silva, Víctor Moreno-González, Eva Trapote,
  Outi Vainio, Antti Vehkaoja, Ken Yoda, Katherine Zacarian, Ari Friedlaender,
  and Christian Rutz.
\newblock A benchmark for computational analysis of animal behavior, using
  animal-borne tags, 2023.

\bibitem{hagiwara2023beans}
Masato Hagiwara, Benjamin Hoffman, Jen-Yu Liu, Maddie Cusimano, Felix
  Effenberger, and Katie Zacarian.
\newblock Beans: The benchmark of animal sounds.
\newblock In {\em ICASSP 2023-2023 IEEE International Conference on Acoustics,
  Speech and Signal Processing (ICASSP)}, pages 1--5. IEEE, 2023.

\bibitem{bermant2021biocppnet}
Peter~C Bermant.
\newblock Biocppnet: Automatic bioacoustic source separation with deep neural
  networks.
\newblock {\em Scientific Reports}, 11(1):23502, 2021.

\bibitem{hagiwara2023aves}
Masato Hagiwara.
\newblock Aves: Animal vocalization encoder based on self-supervision.
\newblock In {\em ICASSP 2023-2023 IEEE International Conference on Acoustics,
  Speech and Signal Processing (ICASSP)}, pages 1--5. IEEE, 2023.

\bibitem{hagiwara2022modeling}
Masato Hagiwara, Maddie Cusimano, and Jen-Yu Liu.
\newblock Modeling animal vocalizations through synthesizers, 2022.

\bibitem{batteau1967man}
Dwight~Wayne Batteau and Peter~R Markey.
\newblock Man/dolphin communication.
\newblock {\em Report on contract}, (00123), 1967.

\bibitem{simard2018mycorrhizal}
Suzanne~W Simard.
\newblock Mycorrhizal networks facilitate tree communication, learning, and
  memory.
\newblock In {\em Memory and learning in plants}, pages 191--213. Springer,
  2018.

\bibitem{bais2004plants}
Harsh~Pal Bais, Sang-Wook Park, Tiffany~L Weir, Ragan~M Callaway, and Jorge~M
  Vivanco.
\newblock How plants communicate using the underground information
  superhighway.
\newblock {\em Trends in plant science}, 9(1):26--32, 2004.

\bibitem{nitrogenTransfer}
Xin-Hua He, Christa Critchley, and Caroline Bledsoe.
\newblock Nitrogen transfer within and between plants through common
  mycorrhizal networks (cmns).
\newblock {\em Critical Reviews in Plant Sciences}, 22(6):531--567, 2003.

\bibitem{gorzelak2015inter}
Monika~A Gorzelak, Amanda~K Asay, Brian~J Pickles, and Suzanne~W Simard.
\newblock Inter-plant communication through mycorrhizal networks mediates
  complex adaptive behaviour in plant communities.
\newblock {\em AoB plants}, 7, 2015.

\bibitem{song2010interplant}
Yuan~Yuan Song, Ren~Sen Zeng, Jian~Feng Xu, Jun Li, Xiang Shen, and
  Woldemariam~Gebrehiwot Yihdego.
\newblock Interplant communication of tomato plants through underground common
  mycorrhizal networks.
\newblock {\em PloS one}, 5(10):e13324, 2010.

\bibitem{barto2012fungal}
E~Kathryn Barto, Jeffrey~D Weidenhamer, Don Cipollini, and Matthias~C Rillig.
\newblock Fungal superhighways: do common mycorrhizal networks enhance below
  ground communication?
\newblock {\em Trends in plant science}, 17(11):633--637, 2012.

\bibitem{ahkami2023emerging}
Amir~H Ahkami, Odeta Qafoku, Tiina Roose, Quanbing Mou, Yi~Lu, Zoe~G Cardon,
  Yuxin Wu, Chunwei Chou, Joshua~B Fisher, Tamas Varga, et~al.
\newblock Emerging sensing, imaging, and computational technologies to scale
  nano-to macroscale rhizosphere dynamics--review and research perspectives.
\newblock {\em Soil Biology and Biochemistry}, page 109253, 2023.

\bibitem{kauffman2021politics}
Craig~M Kauffman and Pamela~L Martin.
\newblock {\em The politics of rights of nature: Strategies for building a more
  sustainable future}.
\newblock MIT Press, 2021.

\bibitem{putzer2022putting}
Alex Putzer, Tineke Lambooy, Ronald Jeurissen, and Eunsu Kim.
\newblock Putting the rights of nature on the map. a quantitative analysis of
  rights of nature initiatives across the world.
\newblock {\em Journal of Maps}, 18(1):89--96, 2022.

\bibitem{rockstrom2024planetary}
Johan Rockstr{\"o}m, Louis Kotz{\'e}, Svetlana Milutinovi{\'c}, Frank Biermann,
  Victor Brovkin, Jonathan Donges, Jonas Ebbesson, Duncan French, Joyeeta
  Gupta, Rakhyun Kim, et~al.
\newblock The planetary commons: A new paradigm for safeguarding
  earth-regulating systems in the anthropocene.
\newblock {\em Proceedings of the National Academy of Sciences},
  121(5):e2301531121, 2024.

\bibitem{favre2009living}
David Favre.
\newblock Living property: a new status for animals within the legal system.
\newblock {\em Marq. L. Rev.}, 93:1021, 2009.

\bibitem{burke2020across}
Anthony Burke and Stefanie Fishel.
\newblock Across species and borders: Political representation, ecological
  democracy and the non-human.
\newblock {\em Non-Human Nature in World Politics: Theory and Practice}, pages
  33--52, 2020.

\bibitem{schwartz2020green}
Roy Schwartz, Jesse Dodge, Noah~A Smith, and Oren Etzioni.
\newblock Green ai.
\newblock {\em Communications of the ACM}, 63(12):54--63, 2020.

\bibitem{verdecchia2023systematic}
Roberto Verdecchia, June Sallou, and Lu{\'\i}s Cruz.
\newblock A systematic review of green ai.
\newblock {\em Wiley Interdisciplinary Reviews: Data Mining and Knowledge
  Discovery}, page e1507, 2023.

\bibitem{carissimo2023limits}
Cesare Carissimo and Marcin Korecki.
\newblock Limits of optimization.
\newblock {\em Minds and Machines}, pages 1--21, 2023.

\end{thebibliography}
\bibliographystyle{unsrt}

\end{document}